\documentstyle[aps,preprint,epsf]{revtex}
\tightenlines
\newcommand{\be}{\begin{eqnarray}}
\newcommand{\ee}{\end{eqnarray}}
\newcommand\del{\partial}

\newcommand\Dels{D \hspace{-0.26cm} /}

\begin{document}


\title{
\begin{flushright} 
{\small BNL-NT-01/2}
\end{flushright} 
Hard Thermal Loops and Beyond in the Finite Temperature
World--Line Formulation of QED} 
\author{R. Venugopalan$^{1,2}$ and J. Wirstam$^2$ 
\\ \mbox{} }

\address{$^1$RIKEN-BNL Research Center, Brookhaven National Laboratory,
Upton NY 11973. \\
$^2$Department of Physics, Brookhaven National Laboratory, Upton, NY 11973. } 
\maketitle

\begin{abstract}
We derive the hard thermal loop action for soft electromagnetic fields
in the finite temperature world--line formulation at imaginary time,
by first integrating out the hard fermion modes from the microscopic
QED action. Further, using the finite $T$ world--line method, we
calculate all static higher order terms in the soft
electromagnetic field. At high $T$, the leading non-linear terms are independent of the
temperature and, except for a term quartic in the time component of
the vector potential, they cancel exactly against the vacuum
contribution. The remaining $T$-dependent non-linear terms become   
more strongly suppressed by the temperature as the number of soft fields increases,
thus making the expansion reliable. Applications of this method to other theories and
problems at the soft scale are also briefly discussed.
    
\end{abstract}

\section{Introduction}
\label{I}

An important feature of interacting field theories at finite temperature is the generation
of thermally induced mass scales, proportional to the coupling constant $g$ \cite{kapusta,bellac}. 
When an ensemble of interacting particles at temperature $T$  
is disturbed, the system responds by creating excitations at the energy 
scale $gT \ll T$. Here we assume $g\ll 1$. Consequently, there appears
another mass scale due to interactions, 
the soft scale $gT$, in addition to the hard scale $T$ that characterizes
the average energy of the constituent particles\footnote{This statement is 
true for QED. In general, there may be additional scales 
that arise due to interactions. In QCD, one also has scale $g^2 T$ due to screening 
of magnetic interactions.}.
Since $(gT)^{-1}$ is much larger than the average particle separation 
$T^{-1}$, it is clear that such an excitation is a collective phenomenon of the system.

A natural question then is how to describe the physics at the soft scale in terms of the
effective degrees of freedom. This problem has been studied intensively during the last decade,
starting with the identification of the relevant Feynman loop diagrams that has to be taken into
account in amplitudes involving soft external fields 
\cite{rob1,braatenrob,frenkeltaylor}.
Named Hard Thermal Loops (HTL), the diagrammatic approach then led to considerations of
the low energy effective action that captures the physics at the soft scale.

Such an
effective action was first considered in \cite{taylorwong}, and a very
simple expression was later obtained in \cite{rob3}. For example,
the behavior of the soft electromagnetic field is described by the following gauge invariant effective Lagrangian,
\be 
L_{{\rm soft}} = - \frac{m_D^2}{2} \, \int \frac{d\Omega}{4\pi} \,
F_{\mu \gamma} \left ( \frac{\hat{K}_{\gamma} \hat{K}_{\sigma}}{(\hat{K}
\cdot \del)^2} \right ) F_{\mu \sigma } \ , \label{photoneff}
\ee
where  $F_{\mu \gamma}$ is the field strength tensor, $\hat{K}_{\mu} = (i, \hat{k})$ with $\hat{k}$
the unit three vector, $\del_{\mu}$ the partial derivative
and $m_D^2 = (e^2T^2/6)$ the thermally induced Debye mass.
The HTL action was later on derived in a variety of approaches, 
for example, from transport theory by truncating the Schwinger-Dyson
equations \cite{blaizot} as well as by purely classical kinetic treatments \cite{kellyetal}.

In this paper, we demonstrate that the HTL effective theory can be obtained
rather straightforwardly from the full microscopic action by simply imposing a momentum
cut-off and integrating out the hard modes. Such a procedure is close to the renormalization
group picture of Wilson (at the one-loop level), although we will neglect all
cut-off dependent terms that are subleading compared to $T$ in the resulting determinant. In this respect,
our approach is similar to the derivation in
\cite{hansetal}. However, we use a completely different method to evaluate the determinant, 
namely the world--line formulation,
wherein the effective action is written as a quantum-mechanical path integral.

At $T=0$, the world--line method has been developed and successfully
applied to a variety of problems for a long time (see \cite{wlreview}
and references therein). In \cite{mckeon,mckeon2,sato} the
formulation was extended to finite temperatures, and the finite $T$ method
was subsequently used in several studies of the effective action
\cite{shovkovy,barton}. One of our main
objectives here is to demonstrate that this formalism has further
feasible and attractive applications at finite temperature. As we show
with the example of the HTL effective action in QED, the finite $T$ 
imaginary time world--line method can be useful even for calculations
of non--static quantities.

In addition to the HTL effective action, we also discuss the higher
order, non-linear terms.  Indeed, one of the true advantages of the
world--line method, compared to diagrammatic calculations, is that it
comprises the sum of all 1-loop diagrams to a given order in the external
field. Using the relative simplicity
of the world--line method, we demonstrate how the static contribution
to the effective action, for an arbitrary number of soft external
photons, can be obtained.  Although these terms beyond quadratic order
are well known to not be of the HTL type, the non-linear interactions
are needed for studies of light--light scattering and photon splitting
(see \cite{gies} and references therein). Such non-linear processes
have by now been observed in laboratory experiments \cite{experiment},
and they may also be important to explain astrophysical effects such
as the observed $\gamma$-ray spectrum from very massive objects
\cite{bursts}. Even though the kinematical region considered here may not
be encountered around, for example, neutron stars, it is nevertheless
interesting to explore the form of the effective action under different
circumstances and obtain a more general and complete picture. In that
sense, our results reported in this paper adds rather nicely to the
earlier investigations in \cite{elmfors,brandtfrenkel,brandtfrenkel2}.

At this point we should mention that we have not been able to generalize
the non-static case beyond the hard thermal loop action. Including
the dynamical information is clearly of outmost importance. Hopefully 
the results reported here can serve as a first step in that direction.

Finally, it should also be noted that the world--line
formalism provides an intuitive connection between classical kinetic
theory \cite{Israeletal} and the underlying quantum
field theory. At finite $T$, this was noted in \cite{tsunami} and
later explored in \cite{minding}. Remarkably, a world--line approach,
analogous to that used to obtain classical kinetic theory, can be
applied to derive an effective action for small-$x$
physics \cite{smallx}.

The paper is organized as follows. In the next section, we recall how
to go from determinants to quantum-mechanical path integrals and
write down the one--loop effective action for soft gauge fields in
finite temperature QED.  In section III, we explicitly derive the HTL
effective action from the one loop effective action. In section IV, we
go beyond the HTL approximation and obtain a general expression for
the effective action to all orders in the static component of the
gauge field. We end in section V with our conclusions
and a discussion of further applications. Details of the computations
are discussed in the appendices.

\section{From determinants to quantum--mechanical path integrals}
\label{II}

The usefulness of rewriting one--loop effective actions in the
language of first--quantized, quantum--mechanical path integrals was
first emphasized by Strassler \cite{strassler}. He showed that the
method was equivalent to the Bern-Kosower rules \cite{bernkosower} for
one-loop $n$-point scattering amplitudes in gauge theories, the latter
having been derived originally within the framework of string
theory. In this section, we will begin by briefly reviewing the work
of Strassler.  For a comprehensive review of later developments, we
refer the reader to \cite{wlreview}.  Following \cite{mckeon,mckeon2},
we then proceed to the finite temperature imaginary time case.

\subsection{Scalar QED at $T=0$}

To be specific, consider as an example scalar QED in four Euclidean dimensions.
The effective action $\Gamma_{{\rm eff}}$ for an electromagnetic background field is,
\be
\Gamma_{{\rm eff}} = - \log \left [ \int \, {\cal D} \phi^{\ast}\, {\cal D} \phi e^{-\int d^4x (
\phi^{\ast}(-D^2)\phi )} \right ] \ ,
\ee
where $D^2 = (\del_{\mu} + ieA_{\mu})^2$, with $A_{\mu}$ the electromagnetic vector field
and $e$ the electric charge. Performing the Gaussian integration, and using the 
Schwinger proper time method, we obtain
\be 
\Gamma_{{\rm eff}} = \log {\rm Det} \left [ -D^2 \right ] = -\int d^4x \, \int_0^{\infty} 
\frac{dt}{t} \langle x | e^{-t(-D^2)} |x\rangle \ . \label{gammascalar}
\ee
We can interpret the matrix element in Eq.\ (\ref{gammascalar}) as a transition 
amplitude, and express it in the form of a quantum mechanical path integral
\be
\Gamma_{{\rm eff}} = -\int_0^{\infty}\frac{dt}{t} \int d^4x \, \, {\cal N} \! 
\int_{\stackrel{{\scriptstyle x_{\mu} (t) =x}}{x_{\mu} (0)=x}} {\cal D}x \, e^{-S_{{\rm w.l.}}} =
-\int_0^{\infty}\frac{dt}{t} \, {\cal N} \! \int_{\stackrel{{\scriptstyle x_{\mu} (t) =}}{x_{\mu} (0)}} 
{\cal D}x \, e^{-S_{{\rm w.l.}}} \ , \label{gammawl}
\ee
with the world-line action $S_{{\rm w.l.}}$ given by the expression
\be
S_{{\rm w.l.}} = \int_0^t d\tau \left [ \frac{[\dot{x}_{\mu}(\tau )]^2}{4} 
-ie \dot{x}(\tau )\cdot A (x(\tau )) \right ] \ . \label{wlaction}
\ee
Here $\dot{x}^{\mu} = dx^{\mu}/d\tau$. The normalization constant 
${\cal N}$ in the effective action ensures that the free ($e=0$) theory satisfies the condition 
\be
{\cal N}\int_{\stackrel{{\scriptstyle x_{\mu} (t) =x}}{x_{\mu} (0)=x}} {\cal D}x \, e^{-\int_0^t d\tau  
(\dot{x}_{\mu}^2/4)} = \frac{1}{16\pi^2 t^2} \ . \label{pinorm}
\ee

Physically, Eq.\ (\ref{gammawl}) can be interpreted as the path
integral for the first--quantized theory of a particle interacting
with a classical background field.  The endpoint positions of the
particle are constrained by $x(0)=x(t) = x$, with an integration over
all possible locations $x$ and intervals $t= \Delta \tau$. It should
be noted though that the variable $\tau$ does not represent any real
time but is merely a parametrization of the trajectory.  As such, a
re--parametrization $\tau \rightarrow \tilde{\tau}(\tau )$ does not
affect the classical action if the associated square-root of the
one-dimensional metric, the einbein $\epsilon$, transforms in the
proper way. However, the complete path integral is not invariant
unless one integrates over the space of all metrics as well \cite{polyakov}. In the
case at hand, the einbein has been fixed, without loss of generality,
to $\epsilon = 2$.

The above method can easily be generalized to include spinning
particles as well. This is achieved by representing the trace over
spinor indices as a path integral over anticommuting Grassmann fields
in a supersymmetric generalization of Eq.\ (\ref{gammawl})
\cite{strassler,brink}. This procedure applies in general to internal
symmetries, and as was shown in \cite{tracegrass}, even additional
internal indices such as color can be represented by a true scalar
world--line action.

\subsection{QED at Finite Temperature}

We will now derive the corresponding form of the effective action at
finite T for soft external photons in QED, where the soft photons have
a momenta $p \sim eT \ll T$. We will further assume that the soft
momenta are much larger than the electron mass, $p\gg m$, so that the
electrons can be considered massless.  The microscopic, finite
temperature imaginary time action is
\be
S_{{\rm micro}}= \int_0^{\beta}dx_4 \, \int d^3x \left [ \frac{1}{4} \tilde{F}_{\mu \nu} \tilde{F}_{\mu \nu} + i\bar{\tilde{\psi}}\gamma_{\mu}(\del_{\mu} +ie\tilde{A}_{\mu} )  
\tilde{\psi} \right ]  \ , 
\label{qedaction}
\ee
where $x_4 \! \in \! [0, \beta \!= \! 1/T]$ and $\tilde{F}_{\mu \nu} =
\del_{\mu}\tilde{A}_{\nu} - \del_{\nu}\tilde{A}_{\mu}$, with the
tildes for later notational simplicity. As is well known in the
imaginary time formalism, the gauge fields are periodic in the compact
direction, $\tilde{A}_{\mu} (\vec{x},x_4 +\beta) = \tilde{A}_{\mu}
(\vec{x},x_4)$, whereas the fermions are anti--periodic,
$\tilde{\psi}(\vec{x},x_4 +\beta) = -\tilde{\psi}(\vec{x},x_4)$.

Since both the fermion and gauge fields contain all possible momenta, 
we pick a scale $\Lambda$, such that $eT \ll \Lambda \ll T$, and then 
split the fields into two parts; one part contains the soft momenta and the other, to be integrated out, 
contains the hard momenta. 
By performing this cut--off procedure, we introduce an explicit $\Lambda$-dependence into the effective action
and also break the gauge invariance. However, the terms in the resulting effective action
that depend on the scale $\Lambda$ are subleading, since  
$\Lambda /T \ll 1$. They can therefore be ignored here since we will only be interested in the 
leading dependence on $T$. The remaining terms, being independent of $\Lambda$, thus have to be
gauge invariant.

To derive the one-loop effective action 
for a soft electromagnetic field one can neglect the hard photons. We 
therefore take $\tilde{A}_{\mu} = A_{\mu}$. Simultaneously, we write $\tilde{\psi} =\psi +\Psi$, 
where $\Psi$ contains the hard momenta.
We then use the background field method \cite{abbott} to
expand the action in Eq.\ (\ref{qedaction}) to quadratic order in $\Psi$. 
Integrating out the hard field, we obtain the effective action for the soft photon modes
\be
S = \int d^4x \left ( \frac{1}{4}F_{\mu \nu}^2 + i\bar{\psi}\gamma_{\mu}(\del_{\mu} +i
eA_{\mu} )  
\psi \right ) - \log {\rm Det} \left [i \Dels \right] \equiv \tilde{S}_{{\rm soft}}+ 
S_{{\rm soft}}\ , \label{detaction}
\ee
where $\Dels = \gamma_{\mu}(\del_{\mu} +ieA_{\mu} )$ and $S_{{\rm soft}}$ contains the determinant.
The determinant can be re--written as
\be
{\rm Det} \left[ i \Dels \right] = && \sqrt{\left ( {\rm Det} \left[ i\Dels\right ] \right )^2 } = 
\sqrt{ {\rm Det} \left [ -D^2 {\bf 1}+ie\sigma_{\mu \nu} F_{\mu \nu} \right ] } = 
\sqrt{ {\rm Det} \left [ -D^2 \right ] \left ({\bf 1} \!- \! ie\sigma_{\mu \nu} F_{\mu \nu}/\! D^2 \right )} 
\ , \label{rewritedet}
\ee
where $\sigma_{\mu \nu} = [\gamma_{\mu}, \gamma_{\nu} ] /4$. 

Now, $\del_{\mu}$ acting on the hard field gives a contribution $\sim
T$, whereas the magnitude of the soft field $A_{\mu}$ is $\sim T$. By
power counting arguments, it follows that the term $eF_{\mu
\nu}/D^2$ is sub--leading compared to the ${\bf 1}$ in Eq.\
(\ref{rewritedet}) when it comes to the HTL effective action. Thus 
the dominant correction from integrating out the hard fermion field
becomes (after taking a trivial trace over spinor indices)
\be
S_{{\rm soft}} = -2\, {\rm Tr} \log  \left [ -D^2 \right ] \ . \label{actionhardfermion}
\ee

Apart from an overall numerical factor, the important difference between Eq.\ (\ref{actionhardfermion})
and the corresponding equation at $T=0$, Eq.\ (\ref{gammascalar}), stems from the fact that the finite $T$ 
determinant has to respect the appropriate periodicity conditions in the compact $x_4$-direction.
In this case, it therefore has to be evaluated on the space of anti--periodic functions, 
due to the anti--periodic boundary condition of the fermion field. 
As was shown in ref.\cite{mckeon2}, the resulting quantum--mechanical path integral is
\be
S_{{\rm soft}} = 2\sum_{n=-\infty}^{\infty} (-1)^n \int_0^{\infty}\frac{dt}{t} \, {\cal N} \! 
\int_{\stackrel{{\scriptstyle x_{\mu} (t) =}}{x_{\mu}(0)+n\beta\delta_{\mu 4}}} 
\! \! \! \! \! {\cal D}x \, e^{-S_{{\rm w.l.}}} \ , \label{fermionaction}
\ee
where $S_{\rm w.l.}$ is the world--line action in Eq.\ (\ref{wlaction}).

In the imaginary time formalism of world--lines, a finite temperature corresponds to the fact that the path
can wind an arbitrary number of times around the cylinder $S^1 \times R^3$ before returning
to the starting point. For $n=0$ there is no winding, so that particular term is insensitive to the compact form of $x_4$
and therefore gives the $T=0$ contribution.

\section{Hard thermal loops from world-lines}
\label{III}

In this section, we will derive the HTL effective action for soft
photons from Eq.\ (\ref{fermionaction}). Many of the details of the 
computation are given in Appendix A. We begin by noting 
that Eq.\ (\ref{fermionaction}) is the expectation value for the
Wilson line, and can be expanded in powers of $A_{\mu}$,
\be
S_{{\rm soft}}= \left \langle e^{ie\int_0^{\infty}d\tau \dot{x}\cdot A} \right \rangle = 
\left \langle S_{{\rm soft} \, (0)} + S_{{\rm soft} \, (1)} + S_{{\rm soft} \, (2)} + 
\ldots \right \rangle \ , \label{expandaction}
\ee
where the subscript $(i)$ indicates the number of external fields.
The individual contributions are all gauge invariant under the periodic
gauge transformations $A_{\mu} \rightarrow A_{\mu}
-(1/e)\del_{\mu}\omega$, where $\omega (\vec{x}, x_4+n\beta ) = \omega
(\vec{x}, x_4)$.

To lowest order in the coupling constant, the path integral describes a
free theory, and the corresponding contribution to $S_{{\rm soft}}$ is
easily obtained. Performing the substitution \be x_{\mu}(\tau) \rightarrow
u_{\mu}(\tau) + n\beta\delta_{\mu 4} \tau /t + z_{\mu} \ ,
\label{shiftx} \ee with $z_{\mu}$ a $\tau$-independent constant and
$u_{\mu}(t) = u_{\mu}(0) = 0$, we use Eq.\ (\ref{pinorm}) to get 
\be
S_{{\rm soft} \, (0)}= && 2V_4 \sum_{n=-\infty}^{\infty} (-1)^n
\int_0^{\infty} \frac{dt}{t} \left ( \frac{1}{16\pi^2 t^2} \right ) e^{-n^2\beta^2 /4t} = 
\frac{V_4}{8\pi^2}\int_0^{\infty} \frac{dt}{t^3} + \nonumber \\ &&  \frac{V_4}{4\pi^2} 
\sum_{n=1}^{\infty}(-1)^n \int_0^{\infty} \frac{dt}{t^3} e^{-n^2\beta^2 /4t} \ ,
\ee
with $V_4$ the four--volume. The first term is ultraviolet divergent but 
independent of the temperature, whereas the second term gives
\be
\left . S_{{\rm soft}\, (0)} \right |_{T>0} = - \frac{7\pi^2T^4 V_4}{180} \ ,
\ee
which is the expected result for free fermions when the infrared cutoff $\Lambda$ is neglected.

The next term $S_{{\rm soft} \, (1)}$ in Eq.\ (\ref{expandaction}) is linear in the soft external
field and vanishes, due the sum over winding modes being odd and
$\int_0^t d\tau_1 \dot{u}_{\mu}(\tau_1) = u_{\mu}(t)-u_{\mu}(0)
=0$. This result is of course just a manifestation of Furry's theorem.

Now consider the term quadratic in the external field, $S_{{\rm soft} \, (2)}$. For simplicity, we will work 
in a gauge where $A_4$ is time independent, namely $\del_4 A_4 = 0$. The spatial part $A_i$ ($i =1,2,3$) of 
the vector field remains nonstatic and transforms under the residual static gauge transformations as 
$A_i \rightarrow A_i -(1/e)\nabla_i \omega$.  There are three kinds of terms to 
consider: ($A_4$-$A_4$), ($A_i$-$A_j$) and ($A_4$-$A_i$). These will be discussed below, with some of the details of 
the derivation given in Appendix A.

\subsection{The $A_4$-$A_4$ term}
\label{thea4section}

Expanding out the effective action to quadratic order, the most general form of the effective 
action is given by Eq.\ (A.2). For the term containing two powers of $A_4$, Eq.\ (A.2) can be 
simplified to read
\be
&& S_{{\rm soft} \, (2)} (A_4{\scriptstyle -}A_4)= -2e^2 \int \! \frac{d^3p}{(2\pi)^3}
\int_0^{\beta}\! dz_4 A_4^{(1)}A_4^{(2)}\sum_{n=-\infty}^{\infty} (-1)^n 
\int_0^{\infty} \frac{dt}{t} e^{-n^2\beta^2/4t} {\cal N} \int_0^t\! d\tau_1 \int_0^{\tau_1} \! d\tau_2
\times \nonumber \\ && 
\int {\cal D}u\left ( \dot{u}_{4}(\tau_1) +\frac{n\beta}{t} \right )
\left ( \dot{u}_{4}(\tau_2) +\frac{n\beta}{t} \right ) 
\exp\left [-\int_0^td\tau \left \{ \dot{u}^2/4 -i\vec{p} \cdot \vec{u} ( \delta(\tau-\tau_2) -
\delta(\tau-\tau_1) ) \right \} \right ] \ , \nonumber \\ \label{afourstart}
\ee
where $A_4^{(1)} = A_4 (\vec{p})$ and $A_4^{(2)} = A_4 (-\vec{p})$.

Following Strassler \cite{strassler}, we now exponentiate the vector potential under the condition that
only terms linear in each $A_4^{(i)}$ contribute:
\be
\left ( \dot{u}_{4}(\tau_i) +\frac{n\beta}{t} \right ) A_4^{(i)} = \left . e^{
( \dot{u}_{4}(\tau_i) +\frac{n\beta}{t})A_4^{(i)}} \right |_{{\rm linear} \, {\rm in} \,
{\rm each} \, A_4^{(i)}} \ . \label{strasslertrick}
\ee
As discussed in appendix A (see Eqs.\ (A.4)--(A.10)), 
this exponentiation allows one to re--write the exponential in Eq.\ (\ref{afourstart}) in terms of 
one dimensional Green's functions of a free particle constrained to a circle of circumference $t$. 
Subsequently, the path integral can be made Gaussian after a change of variables. Since the Green's 
function and its derivatives are known, one finds, by defining $\gamma = \vec{p}^{\, 2}x(1-x)$ and using
Eqs.\ (A.13) and (A.14), that
the finite temperature part ($n \neq 0$) of the effective action\ (\ref{afourstart}) becomes,
\be
\left . S_{{\rm soft} \, (2)} (A_4{\scriptstyle -}A_4) \right |_{T>0} = && 
-\frac{2e^2}{\pi^2} \int \! \frac{d^3p}{(2\pi)^3} A_4(\vec{p}) A_4(-\vec{p})\int_0^{\beta}\! dz_4 \!
\int_0^1 dx \, (1-x)\sum_{n=1}^{\infty} (-1)^n 
\times \nonumber \\ && \left ( \gamma
K_2 (n\beta\sqrt{\gamma}) - \frac{\sqrt{\gamma}}{n\beta} K_1 (n\beta\sqrt{\gamma}) \left [ 1 - 
\delta (x) \right ] \right ) \ . \label{afourinbessels}
\ee

The leading contribution in Eq.\ (\ref{afourinbessels}) is obtained when $\beta\sqrt{\gamma} \rightarrow 0$.
In this limit, we can use
\be 
\lim_{\beta\sqrt{\gamma} \rightarrow 0} K_2 (n\beta\sqrt{\gamma}) = \frac{2}{n^2\beta^2\gamma} \ \ \ ,
\ \ \ \ \lim_{\beta\sqrt{\gamma} \rightarrow 0} K_1 (n\beta\sqrt{\gamma}) = \frac{1}{n\beta\sqrt{\gamma}} \ ,
\ee
since the sum over the winding modes remains convergent. Employing the identity, 
$\int_0^1 dx (1-x)[1- \delta(x)] =0$, with the usual
convention \cite{mckeon,strassler} for the sign function, $\epsilon(0)=0$, we obtain finally the result
\be
\left . S_{{\rm soft} \, (2)} (A_4{\scriptstyle -}A_4) \right |_{T>0} = && 
-\frac{2e^2T^2}{\pi^2} \int \! \frac{d^3p}{(2\pi)^3} A_4(\vec{p}) A_4(-\vec{p})\int_0^{\beta}\! dz_4 \!
\sum_{n=1}^{\infty} \frac{(-1)^n}{n^2} \nonumber \\ = && \frac{e^2T^2}{6}
\int \! \frac{d^3p}{(2\pi)^3} A_4(\vec{p}) A_4(-\vec{p})\int_0^{\beta}\! dz_4  
= m_D^2 \int d^4z \left [ A_4 (\vec{z}) \right ]^2 \ ,   \label{a4result}
\ee
where $ m_D^2 = e^2T^2/6$ is the Debye mass. In the static limit, this
result is well known to be the only leading contribution
\cite{kapusta,bellac}, and gives rise to a screening of the potential
between two static test charges.

\subsection{The $A_k$-$A_j$ term}

The derivation of this part of the HTL effective action is not as straightforward as the $A_4$-$A_4$ term 
because the spatial components of the vector field also depend on $x_4$, $A_i = A_i (\vec{x}, x_4)$. 
Using Eqs.\ (A.12) and (A.15) in Appendix A, the effective Lagrangian for this term becomes
\be
L_{{\rm soft} \, (2)}(A_k{\scriptstyle -}A_j) = && \frac{e^2}{8\pi^2} \sum_{n=-\infty}^{\infty}
(-1)^n \int_0^{\infty} \frac{dt}{t} e^{-n^2\beta^2 /4t} \int_0^1 dx(1-x) e^{-ip_4n\beta x}
e^{-tP^2x(1-x)} \nonumber \\ \times && \left \{ p_kp_j A_k^{(1)}A_j^{(2)} [1-2x]^2 + \frac{2}{t} 
\vec{A}^{(1)} \! \cdot \! \vec{A}^{(2)} [1-\delta (x)] \right \} \ , \label{efflagrforspatial}
\ee
where the shorthand notation is $A^{(1)} = A(P)$ 
and $A^{(2)} = A(-P)$, with $P^2$ a Euclidean four-vector, $P^2 = p_4^2 + \vec{p}^{\, 2}$
and $p_4 = \omega_l = 2\pi lT$ the external Matsubara frequency.

We now split the effective Lagrangian (\ref{efflagrforspatial}) into a piece
$L^{(1)}$ for the term including the delta function, and another $L^{(2)}$ containing the
rest of the expression. The finite $T$ part of $L^{(1)}$ can be evaluated directly:
\be
\left . \! L_{{\rm soft} \, (2)}^{(1)} \right |_{T>0} = 
&& \left . \! \frac{-e^2}{4\pi^2}\vec{A}^{(1)} \!
\cdot \! \vec{A}^{(2)} \! \! \sum_{n=-\infty}^{\infty} \!
(-1)^n \! \int_0^{\infty} \frac{dt}{t^2} e^{-n^2\beta^2 /4t} \int_0^1 \!dx \,\delta (x) \, 
(1 \! - \! x) e^{-ip_4n\beta x -tP^2x(1-x)} \right |_{T>0} \! \! \!  
\nonumber \\ = && \left ( \frac{e^2T^2}{12} \right ) 
\vec{A}^{(1)} \! \cdot \! \vec{A}^{(2)} \ . \label{deltacontribution}
\ee
This term looks like a magnetic mass-term, but since such a term is forbidden in QED \cite{fradkin} it has to be 
cancelled by the contributions from $L^{(2)}$. In addition, it also violates the residual gauge invariance.

As in \cite{mckeon2}, the remaining terms in $L^{(2)}$ can be
simplified further by first using the substitutions $n\rightarrow -n$
and $y=1-x$. If $f(x,n)$ denotes the part of $L^{(2)}$ that only has
to be summed over $n$ and integrated over $x$ to produce the
Lagrangian, we have $\sum_n \int_0^1 dx \, (1-x)  f(x,n) =\sum_n
\int_0^1 dy \, y f(y,n)$, by using the fact that $p_4$ is a
discrete Matsubara frequency. As in the purely static case, we can then
write the result in terms of Bessel functions of the second kind, to find the finite $T$ contribution  
\be
\left . L_{{\rm soft} \, (2)}^{(2)} \right |_{T>0} &=& \left ( \frac{e^2}{4\pi^2} \right ) 
\int_0^1dx(1-2x)\sum_{n=1}^{\infty}(-1)^n K_0 \left ( n\beta \sqrt{P^2x(1-x)} \, \right ) 
\nonumber \\ &\times &
\left \{ p_kp_j A^{(1)}_kA^{(2)}_j (1-2x) \cos [ np_4\beta x ]  + 2P^2 
\vec{A}^{(1)} \! \cdot \! \vec{A}^{(2)} \int_0^x d\sigma \cos [np_4\beta \sigma ] \right \} \ . \label{asquare1}
\ee
To obtain the second term in the above equation, we performed an integration 
by parts in $x$, and then defined the integral over the auxiliary variable $\sigma$ 
to re--write a sine-function as a cosine.    

We now use the following identity for the infinite sum over winding modes   \cite{integraltable},
\be
\sum_{n=1}^{\infty} (-1)^n \cos[n\kappa ]K_0(n\phi ) = && \frac{1}{2} \left [ \gamma_E +\log \left (
\frac{\phi}{4\pi} \right ) \right ] +\frac{\pi}{2} \sum_{m=0}^{\infty} \left [ \frac{1}
{\sqrt{\phi^2 + [(2m+1)\pi -\kappa]^2}} \right ] + \nonumber \\ &&
\frac{\pi}{2} \sum_{m=0}^{\infty} \left [ \frac{1}{\sqrt{\phi^2 + [(2m+1)\pi +\kappa]^2}} \right ]
-\frac{1}{2}\sum_{m=0}^{\infty}\frac{1}{(m+1)} \ , \label{besselsum}
\ee   
where $\gamma_E$ is Euler's constant. In fact, this identity converts 
the sum over winding modes into
a sum over fermion Matsubara frequences. Keeping only the leading temperature dependence of Eq.\ 
(\ref{besselsum}), we show in Eqs.\ (A.16)--(A.19) that the finite $T$ Lagrangian (\ref{asquare1}) 
becomes
\be
&& \left . L_{{\rm soft} \, (2)}^{(2)} \right |_{T>0} \nonumber \\ && =    
\left ( \frac{e^2}{2\pi^2} \right ) A^{(1)}_kA^{(2)}_j \frac{\del^2}{\del \! p_k \del \! p_j}
T \!\sum_{m=-\infty}^{\infty}\int_0^{\infty} d\theta \, \theta^2 \int_0^1 \frac{dx}{x(1-x)}
\log \left [ P^2x(1-x) +\theta^2 + (k_4 - xp_4)^2 \right ] \ , \nonumber \\ \label{asquare5}
\ee
where $k_4 = (2m+1)\pi T$ plays the role of the fermion Matsubara frequency.

Now, to extract the HTL effective action from Eq.\ (\ref{asquare5}) we note that the
leading contribution is equivalent to a hard thermal loop integral, multiplied by $A^{(1)}_k\!A^{(2)}_j$. 
This can be seen as follows. First take the derivatives with respect to $p_k$ and $p_j$ to get
\be
\left . L_{{\rm soft} \, (2)}^{(2)} \right |_{T>0} =  && \left ( \frac{e^2}{\pi^2} \right )
A^{(1)}_kA^{(2)}_j \left [ T \!\sum_{m=-\infty}^{\infty} \int_0^{\infty} d\theta \, \theta^2 \int_0^1 dx  
\left \{ \frac{\delta_{k j}}{P^2x(1-x) +\theta^2 +(k_4 -xp_4)^2} - \right . \right . \nonumber \\ &&
\left . \left . - \frac{2p_jp_k x(1-x)}
{[P^2x(1-x) +\theta^2 +(k_4 -xp_4)^2]^2} \right \} \right ] \ . \label{asquare6}
\ee
Partially integrating over $\theta$ in the first term in Eq.\ (\ref{asquare6}), and neglecting a
temperature independent UV-divergent constant, we are left with
\begin{equation}
\left . L_{{\rm soft} \, (2)}^{(2)} \right |_{T>0} = \left ( \frac{2e^2}{3\pi^2} \right ) 
A^{(1)}_kA^{(2)}_j T \!\sum_{m=-\infty}^{\infty} \int_0^{\infty} d\theta \, \theta^2 \int_0^1 dx
\left ( \frac{\theta^2 \delta_{k j} - 3p_jp_k x(1-x)}{[P^2x(1-x) +\theta^2 +(k_4 -xp_4)^2]^2}
\right ) \ . \nonumber \\ 
\end{equation}

Since the $\theta$-dependent integrand only contains multiples of $\theta^2$, we can
formally interpret $\theta$ as the magnitude of a three-vector, so that $\theta^2 \rightarrow |\vec{\theta}|^2$ and
$d\theta \theta^2 \rightarrow d^3\theta /(4\pi)$. By adding the terms $A^{(1)}_kA^{(2)}_j [
x(\theta_kp_j +\theta_jp_k) - \theta_kp_j]$, which are all odd in $\vec{\theta}$, and changing the 
integration variable $\vec{\theta} = \vec{k} - x\vec{p}$, we obtain the leading contribution,
\be
\left . L_{{\rm soft} \, (2)}^{(2)} \right |_{T>0} = 4e^2 A^{(1)}_kA^{(2)}_j T \!\sum_{m=-\infty}^{\infty}
\int \frac{d^3 k}{(2\pi)^3} \, \frac{k_k k_j}{K^2(K-P)^2} \ , \label{htlintegral}
\ee
where $K^2 = k_4^2 +\vec{k}^2$. Once the sum has been performed, one can continue $p_4$ to arbitrary Euclidean
values, so that it becomes the analytical continuation of the energy in Minkowski space-time. 
When both $p_4 \ll T$ and $|\vec{p}|\ll T$, as is the case when the fields are soft,
the integral in Eq.\ (\ref{htlintegral}) is well known \cite{bellac}.  
The leading contribution from Eq.\ (\ref{asquare5}) is 
\be
\left . L_{{\rm soft} \, (2)}^{(2)} \right |_{T>0} = -\frac{e^2T^2}{12}\vec{A}(P) \cdot \vec{A}(-P) +
\frac{e^2T^2}{6} \int \frac{d\Omega}{4\pi} \hat{k} \cdot \vec{A}(P) \left [ \frac{ip_4}{\hat{K} \cdot P}
\right ] \hat{k} \cdot \vec{A}(-P)  \ . \label{akajcontribution}
\ee

Thus, as anticipated, 
the first term in (\ref{akajcontribution}) cancels the result
in Eq.\ (\ref{deltacontribution}). The
total effective Lagrangian for the spatial part of the electromagnetic field is then
\be
\left . L_{{\rm soft} \, (2)}(A_k{\scriptstyle -}A_j) \right |_{T>0} =
\frac{e^2T^2}{6} \int \frac{d\Omega}{4\pi} \hat{k} \cdot \vec{A}(P) \left [ \frac{ip_4}{\hat{K} \cdot P}
\right ] \hat{k} \cdot \vec{A}(-P)  \ . \label{akajresult}
\ee
This term is gauge invariant under the residual static gauge transformations allowed by the condition $\del_4 A_4=0$. 

\subsection{The $A_4$-$A_k$ term}

Finally, we have to consider the part of the effective action that contains one spatial component $A_k$
together with one power of $A_4$. Since the $A_4$-field is static, we 
have after a Fourier transformation (see Eq.\ (A.2))
\be
S_{{\rm soft}\, (2)}(A_4 {\scriptstyle -} A_k) = && -2e^2 \int \frac{d^3p}{(2\pi)^3} \sum_n (-1)^n \int_0^{\infty}
\frac{dt}{t}e^{-n^2\beta^2 /4t} \, {\cal N} \! \int {\cal D}u e^{-\int_0^t d\tau \dot{u}^2/4} 
\int_0^t d\tau_1 \int_0^{\tau_1} d\tau_2 \times \nonumber \\ &&
\left ( \dot{u}_4(\tau_1) +\frac{n\beta}{t} \right ) \dot{u}_k (\tau_2)
A_4(\vec{p}) A_k(-\vec{p}) e^{i\vec{p} \cdot (\vec{u}(\tau_1) - \vec{u}(\tau_2))} \ .
\ee
However, the $n\beta /t$-term vanishes since the sum is odd in $n$,
whereas the $\dot{u}_4$-term gives, after an exponentiation in the
same way as before,
\be
&& S_{{\rm soft}\, (2)}(A_4 {\scriptstyle -} A_k) = \nonumber \\ && \, \, \, \, 
\left . -\frac{e^2}{8\pi^2} \int \frac{d^3p}{(2\pi)^3} \sum_n (-1)^n \int_0^{\infty} \! \frac{dt}{t^3}
e^{-n^2\beta^2 /4t} \int_0^t \! d\tau_1 \int_0^{\tau_1} \! d\tau_2 
e^{-\vec{p}^{\, 2} G_B(\tau_1,\tau_2) -i\vec{p} \cdot \vec{A} \del_{\tau_2}G_B(\tau_1,\tau_2)} \right 
|_{\stackrel{{\scriptstyle {\rm linear} \,  {\rm in}}}{A_k \, {\rm and} \, A_4}} \ . \nonumber \\
\ee
But this part of the effective action has no dependence on
$A_4$. Therefore, the exponent cannot be expanded to give any term
linear in $A_4$, and the Lagrangian vanishes to all orders in $p/T$.

\subsection{Summing the contributions} 

The complete effective action to quadratic order in the soft photons
then becomes, by adding the contributions from Eqs.\ (\ref{a4result})
and (\ref{akajresult})
\be
S_{{\rm soft}} = m_D^2 \int d^4x \left ( \left [ A_4(\vec{x}) \right ]^2 + \int \frac{d\Omega}{4\pi} \hat{k} 
\cdot \vec{A}(x) \left \{  \frac{i\del_4}{\hat{K} \cdot \del} \right \}  \hat{k} \cdot \vec{A}(x)  \right ) \ . 
\label{finalresult}
\ee
We have thus derived the effective action by first integrating out the
hard modes from the microscopic action, and then evaluating the
determinant form of the one-loop effective action in Eq.\
(\ref{detaction}) as a quantum-mechanical path integral. The general
HTL action in Eq.\ (\ref{photoneff}) of course reduces to the above
result when the gauge condition $\del_4 A_4=0$ is imposed.

\section{Beyond hard thermal loops}
\label{IV}

In this section, we turn to the higher order terms in the effective
action. Although these terms do not have HTL's, it is nevertheless
interesting to study the form of the effective action in the high $T$ phase, 
and obtain a systematic expansion in powers of $p/T$. As mentioned
already in the introduction, the non-linear effective action is a
useful tool in astrophysical applications \cite{elmfors}.  Moreover,
one could use the resulting non-linear expression to contract and
thermalize any pair of external legs, and by this procedure obtain
higher corrections to, for example, the pressure.

The power of the world--line approach becomes manifest as one goes to
higher orders. For instance, to obtain the dominant term at high
temperature, one should note that individual loops in the diagrammatic
approach may contain superficially leading terms that only cancel when
all contributing graphs are summed. In marked contrast, within the world--line formalism, all diagrams are accounted for in the effective
action.

In \cite{brandtfrenkel,brandtfrenkel2} a diagrammatic approach was
used to argue that all terms with $N \geq 4$ external photons have a
finite limit when $T\rightarrow \infty$. In the
world-line formulation, we will verify explicitly that this is true
for the static terms. In principle\footnote{With the operator $D^2$,
this is true for scalar QED whence the boundary conditions in $x_4$ is
changed. In QED, one would also have to take into account the $F_{\mu
\nu}$-term to get the correct factor.}, we can also obtain all
corrections in a power expansion of $p/T$. For $N=4$, parts of our
results overlap with the world--line calculation in \cite{barton}.

Proceeding in the same way as before, by generalizing straightforwardly 
Eq.\ (A.12) of Appendix A to the $N$'th order, we have in the static case
\be
S_{{\rm soft} \, (N)} = && \frac{(-ie)^N}{8\pi^2} \int_0^{\beta} dx_4 \int \frac{d^3p_1}{(2\pi )^3} \! \cdots \!
\frac{d^3p_N}{(2\pi )^3} \delta^3 (p_1 + \ldots +p_N) \sum_n (-1)^n \int_0^{\infty} dt t^{N-3} 
e^{-n^2\beta^2 /4t} \times \nonumber \\  
&& \! \exp \left [ \frac{n\beta}{t}\sum_{i=1}^N \! A_4^{(i)} \right ] \int_0^1 dx_1 \! \cdots \!
\int_0^{x_{N-1}} dx_N \exp \left [ 
-\sum_{\stackrel{{\scriptstyle i,j=1}}{i<j}}^N
\left \{ t\vec{p}_i\cdot \vec{p}_j(x_i-x_j)(1-(x_i-x_j)) \right \} \right . \nonumber \\ && \! \left .
-\sum_{\stackrel{{\scriptstyle i,j=1}}{i<j}}^N \left \{
i(\vec{p}_i \cdot \vec{A}^{(j)} - \vec{p}_j \cdot \vec{A}^{(i)})(1-
2(x_i -x_j)) +\frac{2}{t} A^{(i)} \cdot A^{(j)} (1-\delta (x_i -x_j)) \right \} \right ] \ , \nonumber \\ \label{sforgeneraln}
\ee
where $A^{(i)} = A(p_i)$, and the only term in the field--dependent
exponential to be kept is the one linear in each $A^{(i)}$. Since the
effective action reflects the symmetries of the original microscopic
action, and in particular the invariance under charge conjugation, all
odd powers of the vector field vanish and the only contribution comes
from $N$ even. This generalizes the explicit calculation for a single
external field discussed in section III.

We now expand the exponential $\exp (n\beta \sum_{i=1}^N A_4^{(i)}/t)$
in powers of $A_4$. For odd powers, the sum over $n$ is odd and
vanishes. Hence we only have to consider even powers. The
effective action could still contain an arbitrary even number $k$
($0\leq k \leq N$) of powers of $A_4$, though.  Following the general
procedure outlined in appendix A, we then have
\begin{equation}
S_{{\rm soft}\, (N)} = \frac{(-ie)^N}{8\pi^2} \! \int_0^{\beta} \! dx_4 \int \! \frac{d^3p_1}{(2\pi )^3} \! \cdots \!
\frac{d^3p_N}{(2\pi )^3} \delta^3 (p_1 \!+ \! \ldots \!+ \! p_N) \int_0^1 \! dx_1 \! \cdots \! \int_0^{x_{N-1}} \! dx_N
\! \sum_{\stackrel{{\scriptstyle k=0}}{k \, {\rm even}}}^{N} \! I_N^{(k)} \ , \label{actionwithin}
\end{equation}
where the function $I_N^{(k)}$ consists of three different terms,
\be
I_N^{(k)}= I_{N\, (1)}^{(k)} +I_{N\, (2)}^{(k)} +I_{N\, (3)}^{(k)}  = \sum_{n=-\infty}^{\infty}(-1)^n\left [I_{N\, (1)}^{(k,n)} +I_{N\, (2)}^{(k,n)} +I_{N\, (3)}^{(k,n)} \right ]\label{splitinto3}
\ee
The three terms in Eq.\ (\ref{splitinto3}) correspond to the different ways of expanding the 
exponential in the last line of Eq.\ (\ref{sforgeneraln}) to get $N$ external fields in total, given that the
term $\exp (n\beta \sum_{i=1}^N A_4^{(i)}/t)$ is expanded to an arbitrary even order $k$.
When only the first $p$--dependent term of the last exponential in Eq.\ (\ref{sforgeneraln}) 
contributes with the remaining $(N-k)$ powers of the vector field, one finds,
\be
I_{N\, (1)}^{(k,n)} && = 
\tilde{p}^{(2+k\!-\!N)} (n\beta )^{N-2} K_{2+k\!-\!N} \!
\left ( n\beta \tilde{p} \right ) f_1 (\{x \} ) 
{\cal P}\! \left [ (p_{i_{k+1}} \! \cdots \!  p_{i_N} )( A_{4_1} \!
\cdots \! A_{4_k} )(A_{i_{k+1}} \! \cdots \! A_{i_N}) \right ] , \label{thefirsti}
\ee
and if the $(N-k)$ powers only come from the $p$-independent term
of the exponential, the result is
\be
I_{N\, (2)}^{(k,n)} && = \tilde{p}^{2-(N-k)/2} (n\beta )^{(N+k)/2 -2}
K_{2+(k-N)/2}\left ( n\beta \tilde{p} \right ) f_2 (\{x \} ) \\ &&
 \times {\cal P} \! \left [ ( A_{4_1} \! 
\cdots \! A_{4_k} )[(A^{(k+1)}\cdot A^{(k+2)}) \! \cdots \! (A^{(N-1)}\cdot A^{(N)})] \right ] . \label{thesecondi} 
\ee
Finally, there is also the possibility to expand both terms, giving
\be
I_{N\, (3)}^{(k,n)} && = \theta [(N-4)-k] \!\! \! \sum_{l=1}^{(N-k)/2-1} \! \! 
\tilde{p}^{(2+k+l-N)} (n\beta )^{N-l-2} K_{2+k+l-N}\left ( n\beta \tilde{p} \right )f_3^{(l)} (\{x \} )
\times \nonumber \\ && \! {\cal P}\! \left [ (p_{i_{k+2l+1}} \! \cdots \!  p_{i_N} )( A_{4_1} \!\cdots \! A_{4_k} )  
(A_{i_{k+2l+1}} \! \cdots \! A_{i_N})[(A^{(k+1)}\cdot A^{(k+2)}) \! \cdots \! 
(A^{(k+2l-1)}\cdot A^{(k+2l)})] \right ] \ , \nonumber \\ \label{thethirdi}
\ee
where the $\theta$--function ensures that the term vanishes when $k>(N-4)$.
In all the equations above we used
\be
\tilde{p}^2 = \sum_{\stackrel{{\scriptstyle i,j=1}}{i<j}}^N
\left \{ \vec{p}_i\cdot \vec{p}_j(x_i-x_j)(1-(x_i-x_j)) \right \} \ , \label{definetildep}
\ee
and the functions $f_i(\{x\})$ comprise the dependence on the dimensionless variables $x_i$, together 
with some irrelevant combinatorical factors. ${\cal P}$ denotes both all permutations of 
choosing $k$ factors of $A_4$ and the other $(N-k)$ factors out of the $N$ total, 
as well as all possible ways of contracting the different vector fields with themselves and/or the momentum.

To study the high temperature behavior of the effective action, we
isolate the dependence on $T$ and $p$ in $I_N^{(k)}$. From Eqs.\
(\ref{thefirsti})-(\ref{thethirdi}) we find that all three terms give
different contributions.  As shown in appendix B, the $p$ and $T$
dependence of $I_{N\, (1)}^{(k)}$ becomes
\be
&& I_{N\, (1)}^{(k)} \propto f_1(\{x\})( A_{4_1} \! \cdots \! A_{4_k} )(A_{i_{k+1}} \! \cdots \! A_{i_N}) \times \nonumber \\ &&
\left ( \frac{1}{\tilde{p}} \right )^{N-4} \left [ \int_{-i\infty +\epsilon}^{i\infty +\epsilon}\frac{dz}{2\pi i}
z \Gamma (z) (1-2^{1-z})\zeta (z) \, (-z-1)\cdots (-z-N+3)\left (\beta \tilde{p} \right )^{-z} \times \right . \nonumber \\ &&
\left \{  \delta_{kN} \! \left [ \frac{2\Gamma[(N-4+z)/2]}{\Gamma[(N+z-3)/2]} - 
\frac{\Gamma[(N+z-2)/2]}{\Gamma[(N+z-1)/2]} \right ] +\delta_{k(N-2)} p_{i_{N-1}}p_{i_N} \tilde{p}^{-2} 
\frac{\Gamma[(N+z-2)/2]}{\Gamma[(N+z-1)/2]} + \right . \nonumber \\ && \left . \left .
\sum_{\stackrel{{\scriptstyle j=0}}{j\, {\rm even}}}^{N-4} 2^{N-j-3}\delta_{kj} p_{i_{j+1}}\! 
\cdots p_{i_N} \tilde{p}^{j-N}
\left [ \frac{\Gamma[(j+z)/2]}{\Gamma[(j+z+1)/2]} + \! \! \!
\sum_{m=1}^{(N-j-2)/2} \! \! a_m^{(1)} \frac{\Gamma[(j+z+2m)/2]}{\Gamma[(j+z+2m\!+\!1)/2]} \! \right ] 
\right \} \right ] , \nonumber \\ \label{firstbullet}
\ee
where $\zeta (z)$ is the Riemann zeta-function, $a_m^{(1)}$ contains some irrelevant combinatorical factors and
the sum over $m$ runs up to the largest integer $\leq (N-j-2)/2$. 
Similarly, one can show that the expression for the second term 
$I_{N\, (2)}^{(k)}$, valid for all
$k$ if $N\geq 6$ and for $k\geq 2$ when $N=4$, can be written as  
\be
&& I_{N\, (2)}^{(k)} \propto f_2(\{ x\})( A_{4_1} \! \cdots \! A_{4_k} )
[(A^{(k+1)}\cdot A^{(k+2)}) \! \cdots \! (A^{(N-1)}\cdot A^{(N)})]
\times \nonumber \\ && \left ( \frac{1}{\tilde{p}} \right )^{N-4} \left [ \int_{-i\infty +\epsilon}^{i\infty +\epsilon}
\frac{dz}{2\pi i}z \Gamma (z) (1-2^{1-z})\zeta (z) \, (-z-1)\cdots (-z-(N+k)/2+3)\left (\beta \tilde{p} \right )^{-z} \times 
\right . \nonumber \\ && \left \{  \delta_{kN} \left [ \frac{2\Gamma[(N-4+z)/2]}{\Gamma[(N+z-3)/2]} - 
\frac{\Gamma[(N+z-2)/2]}{\Gamma[(N+z-1)/2]} \right ] +\delta_{k(N-2)} \frac{\Gamma[(N+z-4)/2]}{\Gamma[(N+z-3)/2]} + 
\right . \nonumber \\ && \left . \left .
\sum_{\stackrel{{\scriptstyle j=0}}{j\, {\rm even}}}^{N-4} 2^{(N-j)/2-3}\delta_{kj} 
\left [ \frac{\Gamma[(j+z)/2]}{\Gamma[(j+z+1)/2]} + \! \! \!
\sum_{m=1}^{((N-j)/2-2)/2} \! a_m^{(2)} \frac{\Gamma[(j+z+2m)/2]}{\Gamma[(j+z+2m+1)/2]} \right ] \right \} \right ] 
\ . \label{middlebullet}
\ee
When $N=4$ and $k=0$ it is easier to sum the Bessel function $K_0 (n\beta \tilde{p})$ in Eq.\ 
(\ref{thesecondi}) directly. The leading high temperature term contains a logarithmic dependence
on $T$, but when integrated over $x_i$ this
contribution vanishes. This result, with four external fields, agrees with the result in \cite{barton}.  
Finally, the last term $I_{N\, (3)}^{(k)}$ becomes, for a given $k\leq N-4$,
\be
&& \! \! I_{N\, (3)}^{(k)} \! \propto \! \! \! \sum_{l=1}^{(N-k)/2-1} \! \! \! f_3^{(l)}(\{ x\})( A_{4_1} \! \cdots \! A_{4_k} ) 
(A_{i_{k+2l+1}} \! \cdots \! A_{i_N}) 
[(A^{(k+1)}\cdot A^{(k+2)}) \! \cdots \! (A^{(k+2l-1)}\cdot A^{(k+2l)})] \times \nonumber \\ && 
\left ( \frac{1}{\tilde{p}} \right )^{2(N-l)-4-k}\left [ \int_{-i\infty +\epsilon}^{i\infty +\epsilon}
\frac{dz}{2\pi i}z \Gamma (z) (1-2^{1-z})\zeta (z) \, (-z-1)\cdots (-z-(N-l)+3)\left (\beta \tilde{p} \right )^{-z} \times 
\right . \nonumber \\ && \left. \left \{ 2^{N-k-l-3}\frac{\Gamma[(k+z)/2]}{\Gamma[(k+z+1)/2]} + 
\sum_{m=1}^{(N-k-l-2)/2} \! a_m^{(3)} \frac{\Gamma[(k+z+2m)/2]}{\Gamma[(k+z+2m+1)/2]} \right \} \right ] \ . \label{lastbullet}
\ee

Even though the Eqs.\ (\ref{firstbullet})-(\ref{lastbullet}) look 
complicated, the analytical structure of the integrals 
allow some simple but powerful conclusions. This is due to the fact that 
the temperature dependence is only contained in the 
term $(\beta {\tilde p})^{-z}$, and in each of the terms in Eqs.\ 
(\ref{firstbullet})-(\ref{lastbullet}) the contour can be closed in
the left side of the complex $z$-plane and evaluated by the residue
theorem. The analytical structure of singularities in 
the complex $z$--plane therefore determines the temperature dependence of 
$S_{{\rm soft}\, (N)}$.

The common function $g(z)=z\Gamma(z)(1-2^{1-z})\zeta (z)$ has
simple poles at $z=-(2n+1)$, with $n$ any non-negative integer. Further, the other
$\Gamma$-functions that enter the expressions always appear in
specific ratios, so that the zeros of the denominators in these ratios 
exactly cancel the
poles of $g(z)$ when the $\Gamma$-function in the numerator
starts to have poles along the negative real axis. This holds
irrespective of the additional factors $[(-z-1)(-z-2)\cdots ]$ in the
numerators that actually cancel several of the poles.
Consequently, there are only simple poles to evaluate. If there were
any higher order poles, the factor $(\beta \tilde{p})^{-z}$ would
contribute with a logarithmic factor, but since this is not the case,
the effective action contains only powers of $p/T$.

Although the above equations give the complete, general formula, a simpler
result can be used for the terms independent of the external momenta.  
This limit corresponds to taking the vector field to be constant, 
and we then have for the effective action,
\be
S = 2\sum_n (-1)^n \int_0^{\infty}\frac{dt}{t} {\cal N} \int_{\stackrel{{\scriptstyle x_{\mu} (t) =}}
{x_{\mu}(0)+n\beta\delta_{\mu 4}}} \! \! \! \! \! {\cal D}x e^{-\int_0^t d\tau \dot{x}^2/4} e^{ieA_{\mu}\int_0^t d\tau 
\dot{x}_{\mu}} = \frac{2T^4V_4}{\pi^2} \sum_n \frac{(-1)^n}{n^4} e^{ien\beta A_4} \ ,
\ee
giving for the finite $T$ effective potential $V(A_4)$,
\be
V(A_4) = \frac{4T^4}{\pi^2} \sum_{n=1}^{\infty} \frac{(-1)^n}{n^4} \cos (en\beta A_4) \ .
\ee
Thus the potential is periodic in $A_4$, with period $2\pi T/e$, as was first noted in \cite{oldqcdreview}. 
Restricting $A_4$ to the interval $[-\pi T/e, \pi T/e]$
and writing $A_4 = \pi Tq/e$, $-1\geq q \leq 1$, we have
\be
V(A_4) = -\frac{2T^4\pi^2}{3} \left ( \frac{1}{45} - \frac{1}{24} \left [ 1-(q_m -1)^2 \right ]^2 \right ) \ ,
\ee
where $q_m = (q+1)_{{\rm mod} 2}$ (see \cite{smilga} for a discussion on this effective potential). 
Expanding the potential gives,
\be 
V(A_4) = -\frac{7\pi^2T^4}{180} +m_D^2 (A_4)^2 - \frac{e^4}{12\pi^2}(A_4)^4 
\ . \label{a4potential}
\ee 
This result agrees with the dimensionally reduced theory in
\cite{landsman}\footnote{For QCD, results for the one
loop static potential were first derived by Nadkarni\cite{Nadkarni}.}, 
once the variables are properly rescaled to the canonical mass dimensions in 4d.
The result for $V(A_4)$ can of course be obtained from 
Eqs.\ (\ref{firstbullet})-(\ref{lastbullet}) as well, and we have checked that this is indeed the case.
More importantly though, they also give all the additional derivative terms.

Apart from the quartic term in Eq.\ (\ref{a4potential}), there is an infinite sequence, with different $N$, of leading
terms that are independent of the temperature when $p/T\rightarrow 0$. From Eqs.\
(\ref{firstbullet})-(\ref{lastbullet}) the temperature independent
terms must come from the poles at $z=0$, corresponding to the $k=0$
terms. Since the $k=0$ terms are the only ones that also appear at
$T=0$, it is interesting to compare the two contributions.  We find
from computing the $T=0$ contribution of Eq.\ (\ref{actionwithin}),
arising from the $n=0$ term in Eq.\ (\ref{splitinto3}), that all
$T$-independent terms in Eqs.\ (\ref{firstbullet})--(\ref{lastbullet})
are completely cancelled. 

It is straightforward to also include the $F_{\mu \nu}$ term from Eq.\
(\ref{rewritedet}) into this calculation. In the path integral
formalism \cite{mckeon2}, this term will augment the world--line form
of the effective action by a factor proportional to
 \be
{\rm Tr} \exp \left [ 2\sum_{i=1}^N \sigma_{k \nu} p_{k_i}A_{\nu}^{(i)} \right ] \ ,
\label{fmunu1}
\ee
where $k=1,2,3$ and the trace is over spinor indices.  Now, this term
occurs with the same powers of the variables $A_{\mu}$, $p$, $t$ and
$n$ as the second term in the last exponential of Eq.\
(\ref{sforgeneraln}), namely the term 
\be \exp \left
[ \sum_{\stackrel{{\scriptstyle i,j=1}}{i<j}}^N \left \{ i(\vec{p}_i
\cdot \vec{A}^{(j)} - \vec{p}_j \cdot \vec{A}^{(i)})(1- 2(x_i -x_j))
\right \} \right ] \ .  
\label{fmunu2}
\ee 
The structural dependence on these
variables is the crucial ingredient to prove that a cancellation
indeed occurs. Since we know that all the $T$-independent terms in
Eq.\ (\ref{sforgeneraln}) are cancelled against the $T=0$ effective
action, the identical structures of Eq.\ (\ref{fmunu1}) and 
Eq.\ (\ref{fmunu2}) ensure that this holds true also for the 
$F_{\mu \nu}$
part. Consequently, apart from the quartic term in Eq.\
(\ref{a4potential}), all other higher order terms in the effective
action for soft photons in QED are suppressed by factors of $T$.

As can be seen from Eqs.\ (\ref{firstbullet})-(\ref{lastbullet}), the polynomial
in $[(-z-1)(-z-2)\cdots]$ cancels all of the poles up to order $N$, with the exception
of the pole at $z=0$. Naively, one could imagine that the high temperature expansion would yield
a series like $(1+\beta^2p^2 +\beta^4p^4 + \ldots )$, where the first term corresponds to the 
leading temperature independent part and $p^2$ denotes some generic combination of the external
momenta. Instead, the leading, temperature dependent, contribution for $N$ external fields
is suppressed by $\beta^a$, where $a$ is of the order of $N$. As a result, the
terms in the effective action tend to be more strongly suppressed by powers of the 
inverse temperature as the number of external fields, $N$,
increases. Since the $T$-independent terms by dimensional arguments 
have the schematic form $\del^{(4-N)}(eA)^N$, they are all of the same order for the soft 
fields, where $\del \sim eT$ and $A\sim T$: $\del^{(4-N)}(eA)^N \sim (eT)^{(4-N)}(eT)^N \sim (eT)^4$.
However, the surviving terms that are suppressed by powers of the inverse temperature become smaller
in magnitude as the number of the soft, external fields increases. The expansion in powers of the
soft field is therefore reliable, with the higher orders becoming more and more negligible.           

In \cite{brandtfrenkel} a cancellation of all non-linear terms was
verified in the long wave-length limit, $|\vec{p}|\rightarrow 0$, $p_0
\neq 0$. For slowly varying fields, a similar result was found by
expanding in powers of $m/T$ \cite{elmfors}, $m$ being the electron
mass.  We arrive at the same conclusion in the static
ultrarelativistic limit, apart from the appearance of an
$(A_4)^4$-term, by a direct calculation of all $N\geq 4$ terms.  This
calculation complements the earlier ones and strongly suggests that the $T$
independent non-linear terms beyond quartic order always cancel out,
regardless of the momenta of the external fields. Whether or not the remaining terms are 
more strongly suppressed by the temperature, as in the static case, is not clear at the
moment, although we expect this to be the case due to gauge invariance.

\section{Conclusions}
\label{V}

In this paper, we derived the hard thermal loop effective action for a 
soft electromagnetic field directly from the microscopic, one loop effective 
action in QED. This result was obtained by first
integrating out the hard fermions from the original 
microscopic action, and  subsequently computing the resulting determinant
in the quantum mechanical world--line formalism. 
In addition, we have shown how all leading higher order static terms can 
be obtained with the same method. 
When $T\rightarrow \infty$, all the remaining terms, with the 
sole exception of the quartic term in the $A_4$-potential, 
are cancelled by the $T=0$ terms in the effective action. The remaining terms are generally
suppressed by a factor $\sim \beta^a$, where $a$ is of the order of the number of external fields. 
We would now like to comment on our results and 
speculate about some further applications
wherein the methods described in this paper may prove useful.

First of all, we have neglected all dependence on the momentum cut--off, 
$\Lambda$. This is reasonable as long as we are
only interested in the leading behavior. In general, 
the effective action should of course also depend on the cut--off 
that separates the hard modes from the soft ones. 
A simple momentum cut--off does not respect gauge invariance 
but, in principle,
one could derive the effective action in a gauge where there are 
only physical degrees of freedom present. In this case, 
the momentum cut--off does not pose any problems and one would have a 
true Wilson effective action $S(T,\Lambda)$
for the soft 
electromagnetic fields in the QED plasma, albeit valid only in that 
physical gauge. Such a formulation of the HTL effective
action could be advantageous both for non--static quantities 
as well as for computations of the pressure and entropy in an HTL 
resummation scheme \cite{htlpressure1,htlpressure2}. 
Furthermore, for a quantity like the pressure, even the 
static higher order terms can be used to obtain 
sub--leading contributions \cite{dimred1,dimred2}.

An interesting point to consider further is the connection to the random phase
approximation (RPA). The computation of the one--loop effective action
implies that the corresponding Green's function is obtained in an RPA--in the 
diagrammatic picture this corresponds to resumming a chain of one--loop
fermion bubbles. This corresponds precisely to the derivation of the
dressed propagator in the hard thermal loop approximation
scheme. Thus, in our derivation here, 
we take the opposite approach to the
original HTL derivation; we first calculate the effective action and
from that result deduce the Green's function. The resemblance between the RPA and hard thermal loop approximation 
also suggests a natural direction for  
improving the effective theory further, and hopefully the world--line 
method will prove to be useful in this case as well.

Another interesting extension would be to include also the effects of a 
finite density in our results. This would first of all
generalize the derivation of the HTL effective action. 
Secondly, it would be important to study to what extent  
the non--linear terms cancel also at finite densities, 
since this situation is more likely to resemble situations 
of astrophysical interest.

Finally, a future topic is how to 
include the non--static higher order terms. 
Even though this seems to be a formidable task in general, 
obtaining just the next order correction would be of great interest. 
In particular, a non--Abelian generalization could provide some 
insights into the expansion of the Wilson line in real
time \cite{wilsonlines,rvjw}. 
Such a continuation is not merely of academic interest, but could in fact
have important consequences for the 
observables of the heavy ion collisions at RHIC \cite{wilsonparticles}. 

\begin{center}
{\bf Acknowledgments}
\end{center}
We thank F.\ Gelis and R.\ D.\ Pisarski for useful discussions and 
for reading the manuscript. We also thank L.\ McLerran for emphasizing
the physical context of this work.
R. V. was supported by RIKEN-BNL and by BNL under DOE grant DE-AC02-98CH10886. 
The work of J.W. was supported by The Swedish Foundation for International Cooperation in Research 
and Higher Education (STINT) under contract no 99/665. 

\appendix

\section{Some useful identities}
\renewcommand{\theequation}{A.\arabic{equation}}
\setcounter{equation}{0}

In this appendix we give some of the intermediate stages in the calculation of the hard thermal loop
effective action. With the Fourier expansion
\be 
A_{\mu} (\vec{x}, x_4) = T\sum_m \int\frac{d^3p}{(2\pi )^3}
A_{\mu}(\omega_m, \vec{p}) e^{-i(\omega_mx_4 - \vec{p}\cdot \vec{x})} \ , \label{b1}
\ee
where $\omega_m = 2m\pi T$ is the Matsubara frequency, and using the change of variables in Eq.\ (\ref{shiftx}),
we find the quadratic term in Eq.\ (\ref{expandaction}) to be, before any gauge-fixing, 
\be
&& S_{{\rm soft}\, (2)}= -2e^2 T\sum_m T\sum_l \int\frac{d^3p}{(2\pi )^3} \int\frac{d^3q}{(2\pi )^3} 
A_{\mu}(\omega_m, \vec{p}) A_{\nu}(\omega_l, \vec{q}) \int_0^{\beta}dz_4 \int d^3z 
\sum_{n=-\infty}^{\infty} (-1)^n \times \nonumber \\ && 
\int_0^{\infty} \frac{dt}{t} e^{-n^2\beta^2/4t} {\cal N} \int_0^td\tau_1 \int_0^{\tau_1}d\tau_2 
\int {\cal D}u\left ( \dot{u}_{\mu}(\tau_1) +\frac{n\beta\delta_{\mu 4}}{t} \right ) 
\left ( \dot{u}_{\nu}(\tau_2) +\frac{n\beta\delta_{\nu 4}}{t} \right ) e^{-\int_0^td\tau (\dot{u}^2/4 )}
\times \nonumber \\ && 
\exp \! \left \{i\vec{p} \cdot \left [\vec{u}(\tau_1) +\vec{z}\right ] + i\vec{q} \cdot \left [ \vec{u}(\tau_2) 
+\vec{z}\right ]-i\omega_m\left [ u_4(\tau_1) \! +z_4 + \! \frac{n\beta\tau_1}{t} \right ]
-i\omega_l\left [u_4(\tau_2) \!+ z_4 + \! \frac{n\beta\tau_2}{t}\right ] \right \} \! \nonumber
\\ && \, \, \, = -2e^2 T\sum_m\int\frac{d^3p}{(2\pi )^3}A_{\mu}(\omega_m, \vec{p})A_{\nu}(-\omega_m, -\vec{p})
\sum_{n=-\infty}^{\infty} (-1)^n \int_0^{\infty} \frac{dt}{t} e^{-n^2\beta^2/4t} \times \nonumber \\ &&
{\cal N} \int_0^td\tau_1 \int_0^{\tau_1}d\tau_2 \int {\cal D}u\left ( \dot{u}_{\mu}(\tau_1) +\frac{n\beta\delta_{\mu 4}}{t} \right )\left ( \dot{u}_{\nu}(\tau_2) +\frac{n\beta\delta_{\nu 4}}{t} \right ) e^{-\int_0^td\tau (\dot{u}^2/4 )}
\times \nonumber \\ && \exp \left \{-i\omega_m\left [ u_4(\tau_1)-u_4(\tau_2) \right ] +i \vec{p}\cdot \left [ 
\vec{u}(\tau_1)-\vec{u}(\tau_2)\right ] -i\omega_mn\beta \frac{(\tau_1 -\tau_2)}{t} \right \} \ . \label{b2}
\ee
We now use the trivial identity $u(\tau_i) = \int d\tau [ u(\tau )\delta (\tau-\tau_i)]$, as well as the 
generalization of Eq.\ (\ref{strasslertrick}), 
\be
\left ( \dot{u}_{\mu}(\tau_i) +\frac{n\delta_{\mu 4}\beta}{t} \right ) A_{\mu}^{(i)} = \left . e^{
[ \dot{u}_{\mu}(\tau_i) +\frac{n\delta_{\mu 4}\beta}{t}]A_{\mu}^{(i)}} \right |_{{\rm linear} \, {\rm in} \,
{\rm each} \, A_{\mu}^{(i)}} \ ,
\ee
with the notation $A^{(1)} = A(\omega_m, \vec{p})$, $A^{(2)} = A(-\omega_m, -\vec{p})$, to write Eq.\ (\ref{b2}) as
\be
&& S_{{\rm soft}\, (2)}= -2e^2T\sum_m\int\frac{d^3p}{(2\pi )^3}\sum_{n=-\infty}^{\infty} (-1)^n \int_0^{\infty} 
\frac{dt}{t} e^{-n^2\beta^2/4t}\int_0^td\tau_1 \int_0^{\tau_1}d\tau_2 \times \nonumber \\ &&   
\left . \exp \left [ \frac{n\beta (A_4^{(1)}\!+\!A_4^{(2)})}{t} -i\frac{\omega_mn\beta (\tau_1 -\tau_2)}{t}\right ] 
{\cal N}\! \int {\cal D}u 
\exp \left [ - \int_0^t d\tau \left ( \frac{\dot{u}^2}{4} +J_{\mu}u_{\mu} \right ) \right ] 
\right |_{\stackrel{\! {\scriptstyle {\rm linear}\, {\rm in} }}{{\rm each} \, A_{\mu}^{(i)}}} , 
\ee
where the source term $J_{\mu}=(J_k,J_4)$ is given by
\be
J_{k} (\tau )= && -ip_k \left [ \delta (\tau -\tau_1) - \delta (\tau -\tau_2) \right ] -\sum_{l=1}^2 A_k^{(l)} \delta 
 (\tau -\tau_k)\del_{\tau_k} \ , \\  
J_{4}(\tau )= && i\omega_n \left [ \delta (\tau -\tau_1) - \delta (\tau -\tau_2) \right ] -\sum_{l=1}^2 A_4^{(l)} \delta
(\tau -\tau_k)\del_{\tau_k} \ .
\ee
We can then change the variable of integration, 
\be
\tilde{u}_{\mu} = u_{\mu} - 2\int_0^t d\tilde{\tau} G_B(\tau ,\tilde{\tau}) J_{\mu}(\tilde{\tau}) \ ,
\ee
where $G_B(\tau_1,\tau_2)$ is the one-dimensional Green's function on a circle of circumference $t$,
\be
G_B(\tau_1 , \tau_2) = \left | \tau_1 -\tau_2 \right | - \frac{(\tau_1 -\tau_2)^2}{t} \ ,
\ee
and
\be
\del_{\tau_1}G_B(\tau_1,\tau_2) = \epsilon (\tau_1-\tau_2) -\frac{2(\tau_1-\tau_2)}{t} \ \ , \ \ 
\del_{\tau_1}\del_{\tau_2}G_B(\tau_1,\tau_2) = 2\left [ \frac{1}{t} - \delta (\tau_1-\tau_2)
\right ] \ . 
\ee
The resulting path integral then becomes Gaussian, and with the normalization in Eq.\ (\ref{pinorm})
we obtain (with the implicit condition that only terms containing one power of each $A^{(i)}$ is to be kept),
\be
&& S_{{\rm soft}\, (2)}= -\frac{e^2}{8\pi^2} T\sum_m\int\frac{d^3p}{(2\pi )^3} \sum_{n=-\infty}^{\infty} (-1)^n 
\int_0^{\infty}\frac{dt}{t^3} e^{-n^2\beta^2/4t}\int_0^td\tau_1 \int_0^{\tau_1}d\tau_2 \times \nonumber \\ && \, \, \, \,
\exp \left [ -i\frac{\omega_mn\beta (\tau_1 -\tau_2)}{t} +\frac{n\beta (A_4^{(1)}+A_4^{(2)})}{t} \right ]
\exp \left [-\int_0^t d\tau\int_0^t d\tilde{\tau} J_{\mu}(\tau ) G_B(\tau ,\tilde{\tau})J_{\mu}(\tilde{\tau}) \right ] 
\nonumber \\ && \, \, \, \, = -\frac{e^2}{8\pi^2} T\sum_m\int\frac{d^3p}{(2\pi )^3} \sum_{n=-\infty}^{\infty} (-1)^n 
\int_0^{\infty}\frac{dt}{t^3} e^{-n^2\beta^2/4t}\int_0^td\tau_1 \int_0^{\tau_1}d\tau_2 \times \nonumber \\ && \, \, \, \,
\exp \left [ -i\frac{p_4n\beta (\tau_1 -\tau_2)}{t} +\frac{n\beta (A_4^{(1)}+A_4^{(2)})}{t} \right ]
\exp \left [-P^2G_B(\tau_1,\tau_2) +i\del_{\tau_1}G_B(\tau_1,\tau_2) \times \right . \nonumber \\ && \, \, \, \,
\left . \left \{ p_k(A_k^{(1)}+A_k^{(2)}) -p_4
(A_4^{(1)}+A_4^{(2)}) \right \} -A_{\mu}^{(1)}A_{\mu}^{(2)}\del_{\tau_1}\del_{\tau_2} G_B(\tau_1,\tau_2) \right ] \ ,  
\label{baction1}
\ee
where we have put $p_4 =\omega_m$ and $P^2 = \vec{p}^{\, 2}+p_4^2$ in the last expression. By finally using the identity
\be
\int_0^t d\tau_1 \int_0^{\tau_1}d\tau_2 f(\tau_1 -\tau_2) = 
t^2 \int_0^1 dx(1-x)f(xt) \ ,
\ee
we can write Eq.\ (\ref{baction1}) as
\be
&& S_{{\rm soft}\, (2)}= -\frac{e^2}{8\pi^2} T\sum_m\int\frac{d^3p}{(2\pi )^3} \sum_{n=-\infty}^{\infty} (-1)^n
\int_0^{\infty}\frac{dt}{t} \int_0^1dx(1-x)e^{-tP^2x(1-x) -n^2\beta^2/4t} \times \nonumber \\ && 
\exp \left [ \frac{n\beta (A_4^{(1)}\!+\!A_4^{(2)})}{t} -ip_4n\beta x +
i\epsilon_{\mu 4}\left \{ p_{\mu}(A_{\mu}^{(1)}+A_{\mu}^{(2)})\right \} \left (1-2x\right )
-\frac{2A_{\mu}^{(1)}A_{\mu}^{(2)} \{ 1-\delta (x) \}}{t} \right ] \ , \nonumber \\
\label{baction2}
\ee
with $\epsilon_{\mu 4} = -1$ if $\mu = 4$ and $1$ otherwise. 
The dependence on the vector field can then be expanded again, to recover the specific quadratic terms discussed 
in more detail below.
 
\subsection{The $A_4$-$A_4$ part}

Consider first the terms involving the $A_4$ field.
Since we are using the gauge condition $\del_4A_4=0$ to simplify the actual calculation,
Eq.\ (\ref{baction2}) changes in that there is no $p_4$-dependence when the spatial components $A_k = 0$. Instead
of a sum over Matsubara frequencies the integration over $z_4$ is left over; in other words, 
we do a three-dimensional Fourier transform of $A_4$ instead of Eq.\ (\ref{b1}). 
By expanding the $A_4$-dependent exponential,
\be
\left . e^{\left [ n\beta (A_4^{(1)} + A_4^{(2)})/t -  \frac{2}{t}A_4^{(1)}A_4^{(2)} 
 \left ( 1-\delta (x) \right )\right ]} \right |_{\stackrel{\! {\scriptstyle {\rm linear}\, {\rm in} }}
{{\rm each} \, A_4^{(i)}}} =
\left ( \frac{n^2\beta^2}{t^2} - \frac{2( 1-\delta (x))}{t} \right ) A_4^{(1)}A_4^{(2)} \ ,
\ee
and using the following integral representation of the Bessel function of the
second kind,
\be
K_{\nu}(xz) = \frac{z^{\nu}}{2}\int_0^{\infty}\frac{dt}{t^{\nu +1}} \exp \left [ -\frac{x}{2}\left (t+\frac{z^2}{t} \right )
\right ] \ , \label{besselrep}
\ee
we obtain Eq.\ (\ref{afourinbessels}).

\subsection{The $A_k$-$A_j$ part}

Turning now to the spatial parts of the vector field, we first use Eq.\ (\ref{baction2}) again and put $A_4=0$. We then   
use the expansion
\begin{equation}
\left . e^{\left [ ip_k(A_k^{(1)} + A_k^{(2)})(1-2x)
- 2\vec{A}^{(1)}\cdot \vec{A}^{(2)}( 1-\delta (x))/t \right ]} 
\right |_{\stackrel{\! {\scriptstyle {\rm linear}\, {\rm in} }}
{{\rm each} \, A_k^{(i)}}} =
-\left [\delta_{kj}\frac{2( 1-\delta (x))}{t} + 
p_kp_j \left (1-2x \right )^2 \right ] A_k^{(1)}\!A_j^{(2)} , \nonumber \\
\end{equation}
to obtain Eq.\ (\ref{efflagrforspatial}). Using Eq.\ (\ref{besselrep}) and performing a partial integration in $x$,
we then get Eq.\ (\ref{asquare1}). To find the leading $T$-dependence in the effective action,
only the sums over the square-root terms in Eq.\ (\ref{besselsum}) need to be kept, and by changing 
$m \rightarrow -(m+1)$ in the second square-root we have,
\be 
\left . L_{{\rm soft} \, (2)}^{(2)} \right |_{T>0} = && \left ( \frac{e^2}{8\pi} \right )
\int_0^1dx(1-2x)\, T \!\sum_{m=-\infty}^{\infty} \left [ \frac{p_kp_j A^{(1)}_kA^{(2)}_j (1-2x)}
{\sqrt{-p^2x^2 +(P^2 -2k_4p_4)x +k_4^2}} +  \right . \nonumber \\ + && \left .  
2\int_0^x d\sigma \frac{\vec{A}^{(1)} \! \cdot \! \vec{A}^{(2)}}{\sqrt{P^2x(1-x) +[k_4 -\sigma p_4]^2}}
\right ] \ , \label{asquare2}
\ee
where $k_4 = (2m+1)\pi T$. By integrating over $\sigma$ and performing a partial integration over $x$ in
the second term inside the square brackets, Eq.\ (\ref{asquare2}) becomes,
\be
\left . L_{{\rm soft} \, (2)}^{(2)} \right |_{T>0}  && = \left ( \frac{e^2}{8\pi} \right )
T \!\sum_{m=-\infty}^{\infty} \left [ -4\vec{A}^{(1)} \! \cdot \! \vec{A}^{(2)} \left \{ \sqrt{(k_4-p_4)^2} 
\right \} + \right . \nonumber \\ 
&& + \! \int_0^1dx \left ( \left . \frac{p_kp_j A^{(1)}_kA^{(2)}_j (1-2x)^2 + 2x
\vec{A}^{(1)} \! \cdot \! \vec{A}^{(2)} (P^2 -2p^2x -2k_4p_4)}
{\sqrt{-p^2x^2 +(P^2 -2k_4p_4)x +k_4^2}} \right ) \right ] \ . \label{asquare3}
\ee
To proceed from the above equation, we use that 
\be
\int_0^{\infty} \frac{d\theta \, \theta^2}{(\theta^2 +a)^2} = \frac{\pi}{4\sqrt{a}} \ . \label{thetawrite}
\ee
Since we are discarding all terms but the leading $T$ dependence, we can in fact neglect the term $p_kp_j A^{(1)}_kA^{(2)}_j$
in the numerator $p_kp_j A^{(1)}_kA^{(2)}_j (1-2x)^2$. This can be shown by using Eq.\ (\ref{thetawrite}) and performing
the $x$-integration. We can then write Eq.\ (\ref{asquare3}) as,
\be
&& \left . L_{{\rm soft} \, (2)}^{(2)} \right |_{T>0} = \left ( \frac{e^2}{2\pi^2} \right )
T \!\sum_{m=-\infty}^{\infty} \left [ \int_0^{\infty} d\theta \, \theta^2 \left \{ 
\vec{A}^{(1)} \! \cdot \! \vec{A}^{(2)} \left (-\delta (\theta -1)\, \pi \sqrt{(k_4-p_4)^2} -
\right . \right . \right . \nonumber \\ && \left . \left . \left . - \frac{2}{\theta^2 +(k_4-p_4)^2} \right ) +
A^{(1)}_kA^{(2)}_j \frac{\del^2}{\del \! p_k \del \! p_j} \int_0^1 \frac{dx}{x(1-x)} 
\log \left [ P^2x(1-x) +\theta^2 + (k_4 - xp_4)^2 \right ] \right \} \right ] \ . \nonumber \\ \label{asquare4}
\ee
In the first and second term, we write the sum over $m$ as a contour integral and keep only the finite $T$ part.
For the first term the contribution to the contour integral comes from the cut of the square-root, whereas for the
second the contour can be closed and the contribution evaluated by the residue theorem. 
Remarkably, the two contributions cancel each other and we are left with the last term in Eq.\ (\ref{asquare4}).

\section{The higher order contributions}
\label{B}
\renewcommand{\theequation}{B.\arabic{equation}}
\setcounter{equation}{0}

In going from the expressions for $I^{(k)}_{N\, (i)}$ in Eqs.\ (\ref{thefirsti})-(\ref{thethirdi}) to the results in
Eqs.\ (\ref{firstbullet})-(\ref{lastbullet}), we follow the general 
method described in \cite{mckeon}. To be specific, consider the the sum over winding modes in Eq.\ (\ref{splitinto3})
coming from $I^{(k)}_{N\, (1)}$, Eq.\ (\ref{thefirsti}),
\be
\sum_{n=1}^{\infty} (-1)^n n^{N-2}K_{2-N+k} \left ( n\beta\tilde{p} \right ) \ .
\ee
Using a different integral representation for the Bessel function,
\be
K_{\nu}(x) = \int_0^{\infty} dt \cosh (\nu t)e^{-x\cosh (t)} \ ,
\ee
we have
\be
&& \sum_{n=1}^{\infty} (-1)^n n^{N-2}K_{2-N+k} \left ( n\beta\tilde{p} \right ) = \int_0^{\infty} dt \cosh [(2-N+k)t]
\sum_{n=1}^{\infty} (-1)^n n^{N-2}e^{-n\beta\tilde{p}\cosh (t)} \nonumber \\ 
&& = \, \, \, \, 
\int_0^{\infty} dt \frac{\cosh[(2-N+k)t]}{[\tilde{p}\cosh (t)]^{N-2}} \left (\frac{\del}{\del \beta}\right )^{N-2}
\sum_{n=1}^{\infty} (-1)^n e^{-n\beta\tilde{p}\cosh (t)}\nonumber \\ &&
= \, \, \, \,
\frac{-1}{\tilde{p}^{N-2}} \left (\frac{\del}{\del \beta}\right )^{N-2}\int_0^{\infty} dt
\frac{\cosh[(2-N+k)t]}{[\cosh (t)]^{N-2}} 
\left ( \frac{1}{e^{\beta\tilde{p}\cosh (t)} +1} \right ) \ . \label{sumthebessel}
\ee
Now, with the relation
\be
&& \cosh[(2-N+k)t] = \cosh[(N-2-k)t] = 2^{|N-2-k|-1}[\cosh (t)]^{|N-2-k|} +|N-2-k|
\times \nonumber \\ && \, \, \, \, \sum_{m=1}^{M} \!
\frac{(-1)^m}{m}\left ( \matrix{ |N-2-k|-m-1 \cr m-1 }\right )2^{|N-2-k|-2m-1}[\cosh (t)]^{|N-2-k|-2m} \ , \label{coshrel}
\ee
where the sum over $m$ runs up to the largest integer $M\leq |N-2-k|/2$, we have for the different values of $k$,
\be
&& \frac{-1}{\tilde{p}^{N-2}} \left (\frac{\del}{\del \beta}\right )^{N-2}\int_0^{\infty} \frac{dt}
{e^{\beta\tilde{p}\cosh (t)} +1} \left [ \delta_{kN} \left \{ \frac{2[\cosh (t)]^2 -1}{[\cosh (t)]^{N-2}} \right \}
+\delta_{k(N-2)}\left \{ \frac{1}{[\cosh (t)]^{N-2}} \right \} + \right . \nonumber \\ && \, \, \, \, \, \,  
\sum_{j=0}^{N-4} 2^{N-k-3} \delta_{kj} \left \{
\frac{1}{[\cosh (t)]^j} + \! \! \sum_{m=1}^{M} \! \left .
\frac{a_m^{(1)}}{[\cosh (t)]^{j+2m}} \right \} \right ] \ .
\ee
Here $a_m^{(1)}$ captures the numerical factors in Eq.\ (\ref{coshrel}).

To proceed, we now use the Mellin transform,
\be
g(s) = && \int_0^{\infty} dx \, f(x) x^{s-1} \\ 
f(x) = && \frac{1}{2\pi i}\int_{\epsilon-i\infty}^{\epsilon+i\infty}ds g(s) x^{-s} \ ,
\ee
to write
\be
\left ( e^{x}+1 \right )^{-1} = \frac{1}{2\pi i}\int_{\epsilon-i\infty}^{\epsilon+i\infty}ds \Gamma (s)
(1-2^{1-s})\zeta (s) x^{-s} \ . \label{mellinforfermi}
\ee
Using this equation we can perform the $t$-integration (the integral 
over powers of the hyperbolic cosine function gives ratios of 
$\Gamma$--functions\cite{integraltable}), 
and by taking the $(N-2)$ derivatives with respect to
$\beta$ and inserting the the prefactors of $T$ and $p$, we finally obtain Eq.\ (\ref{firstbullet}). 

In scalar QED, the situation is almost identical, with a few notable differences. Firstly, the overall result
has to be multiplied by $(-1/2)$, to correct for the Fermi statistics and the number of degrees of freedom. Secondly,
the sum over the winding modes in Eq.\ (\ref{sumthebessel}) has no factor $(-1)^n$, so instead of $(-1/[e^x +1])$
one has a Bose-Einstein distribution, $(1/[e^x -1])$. The function $g(s)$ in the Mellin transform then changes to
$\Gamma (s) \zeta (s)$, and the vertical contour in the complex $s$-plane must now be along ${\rm Re}(s)=1+\epsilon$.

\end{document}